# A new form of Tsallis distribution based on the probabilistically independent postulate[*]

Du Jiulin[†]

*Department of Physics, School of Science, Tianjin University, Tianjin 300072, China*

**Abstract**

The current form of Tsallis distribution for a Hamiltonian system with an arbitrary potential is found to represent a simple isothermal situation. In this letter, the *q*-exponential of a sum can be applied as the product of the *q*-exponential based on the probabilistically independent postulate employed in nonextensive statistical mechanics. Under this framework, a new form of Tsallis distribution is suggested. It is shown that the new form of Tsallis distribution can supply the statistical description for the nonequilibrium dynamical property of the Hamiltonian system governed by an *arbitrary* potential, and it is found to be one potential statistical distribution for the dark matter.

**Key words**: Nonextensive statistics; Tsallis distribution; Nonequilibrium dynamics; Fokker-Planck equation.

**PACC**: 0520, 0540

[*] Project supported by the National Natural Science Foundation of China (Grant No. 10675088)
[†] E-mail address: jldu@tju.edu.cn

Nonextensive statistical mechanics since it started in 1988 has obtained very wide applications in many interesting scientific fields. In principle, almost all the formulae and the theory using Boltzmann-Gibbs statistics so far could be generalized under this framework [1-4] (for more detail, see http://tsallis.cat.cbpf.br/biblio.htm ). However, the problems such as under what circumstances, e.g. which class of nonextensive systems and under what physical situation, should the nonextensive statistical mechanics be used for their statistical descriptions have been long-standing [5-20]. In particular, the problem at present appears that [16], unexpectedly, the current form of Tsallis distribution, $f \sim [1-(1-q)\beta H]^{\frac{1}{1-q}}$, employed in nonextensive statistical mechanics is found to be only a simple isothermal or thermal equilibrium situation of the Hamiltonian systems governed by *any* potential, whether for long-range or short-range forces, which, of course, is among the domain of Boltzmann-Gibbs statistics. For a general Langevin equation with an *arbitrary* potential, it is found that there is no possible nonequilibrium dynamics that should use the Tsallis distribution for the statistical description.

Theoretically, the Tsallis distribution function employed in the self-gravitating collisionless system was found to be only an isothermal distribution for any $q \neq 1$ [17]. The example was reported recently in the N-body simulation for a self-gravitating system with the result that the Tsallis distribution is inconsistency generally with the dark matter halos except the isothermal parts for the polytropic index $n \to \infty$ [18], in which, unawarely, the Tsallis distribution function employed was actually isothermal one. On the other hand, however, one can apply the Maxwell *q*-distribution, $f \sim [1-(1-q)\beta mv^2/2]^{\frac{1}{1-q}}$, to deal with some nonequilibrium property of the velocity distribution for self-gravitating and plasma systems [19, 20], where the nonextensive parameter $q \neq 1$ is found to be related to the potential function $\varphi$ ($\varphi$ can be any one) and the temperature gradient $\nabla T$ by the formula expression, $(1-q)\nabla \varphi \sim \nabla T$. The results therefore imply clearly that the Maxwell *q*-distribution can be used for the statistical description of the dynamical system governed by an arbitrary potential when it reaches at the nonequilibrium stationary-state. The applications of the Maxwell $q$-distribution have included the examples such as in the nuclear reaction systems [21-24], in the astrophysical systems (see [17] [19] [2] and the references therein), in the plasmas systems [25-30], in the solar wind theory [31-33], in the non-local distributions in the solar and stellar interior [34-36], and in others [37-42, 1-4]. Obviously, the questions need to be replied, why can the Maxwell $q$-distribution be a possible statistical description for the nonequilibrium dynamical system being at the stationary-state but cannot the current form of Tsallis $q$-distribution? Where does the above discrepancy come from? The purpose of this work is to try a new form of the Tsallis distribution on the basis of the probabilistically independent postulate in nonextensive statistical mechanics, which may be as one reasonable scheme to solve the above discrepancy.

Tsallis proposed the $q$-entropy in 1988 as a generalization of the Boltzmann-Gibbs entropy [43], given by

$$S_q = -k \sum_i P_i^q \ln_q P_i, \qquad (1)$$

where $k$ is Boltzmann constant, the set $\{P_i\}$ are the probabilities of the microscopic configurations of the system under investigation, the parameter $q$ is real number different from unity, the $q$-logarithm is defined as

$$\ln_q x \equiv \frac{x^{1-q} - 1}{1 - q} \quad (x > 0; \ln_1 = \ln x), \qquad (2)$$

the inverse function, the $q$-exponential, is

$$\exp_q x \equiv [1 + (1-q)x]^{\frac{1}{1-q}} \quad (\exp_1 x = \exp x), \qquad (3)$$

if $1+(1-q)x > 0$ and by $\exp_q x = 0$ otherwise. Thus the probability of a system at the value $x_i$ reads $P_i \sim \exp_q x_i$, a power-law distribution. When $q=1$, all the formulae return to be Boltzmann-Gibbs statistics.

Nonextensive statistical mechanics is founded on the basis of the $q$-entropy and the probabilistically independent postulate [43]. The so-called probabilistically independent postulate is that, if the probability of a system at the value $x_i$ is $P_i \sim \exp_q x_i$ and at the value $x_j$ is $P_j \sim \exp_q x_j$, and they are probabilistically independent, then the probability at the value $(x_i+x_j)$ is $P_{ij} = P_i P_j \sim (\exp_q x_i)(\exp_q x_j)$.

The $q$-entropy $S_q$ ($q \neq 1$) is *nonextensive*, namely, if a system composed of two probabilistically independent parts $A$ and $B$, i.e., $P_{ij}^{A+B} = P_i^A P_j^B \ \forall(ij)$ (the probabilistically independent postulate), then the Tsallis $q$-entropy of the system is

$$S_q(A+B) = S_q(A) + S_q(B) + (1-q)k^{-1} S_q(A) S_q(B). \tag{4}$$

Clearly, only if $q=1$ is the entropy *extensive*. Under this framework, one leads to the basic form of the Tsallis distribution used so far in the nonextensive statistical mechanics,

$$f \sim \left[1-(1-q)\beta H\right]^{\frac{1}{1-q}}, \tag{5}$$

with the Lagrange parameter, $\beta = 1/kT$, and the Hamiltonian $H$. According to Eq.(5), if the Hamiltonian of a many-body system is $H = \sum_i p_i/2m + \varphi(\{r_i\})$, one often writes the Tsallis distribution as the form [44-48, 1-4],

$$f \sim \left[1-(1-q)\beta \left(\sum_i p_i^2/2m + \varphi(\{r_i\})\right)\right]^{\frac{1}{1-q}}, \tag{6}$$

which, however, actually contravenes the original postulate of the probabilistic independence. In other words, the $q$-exponential of a sum cannot apply *mechanically* the definition Eq.(3), namely,

$$\exp_q \sum_i x_i \neq \left[1+(1-q)\sum_i x_i\right]^{\frac{1}{1-q}}. \tag{7}$$

But, in fact, in terms of the probabilistically independent postulate, we should express the $q$-exponential of a sum as the product of the $q$-exponential, i.e.,

$$P\left(\sum_i x_i\right) = \prod_i P(x_i), \text{ or}$$

$$\exp_q \sum_i x_i = \prod_i \exp_q x_i. \tag{8}$$

Under this framework, the entropy and the energy are *both* nonextensive in the power-law $q$-distribution. Thus, instead of Eq.(6), a new form of Tsallis distribution for the Hamiltonian many-body system is suggested by

$$f \sim [1-(1-q)\beta\varphi(\{r_i\})]^{\frac{1}{1-q}} \prod_i \left[1-(1-q)\beta\frac{p_i^2}{2m_i}\right]^{\frac{1}{1-q}}. \tag{9}$$

Clearly, only if taking $q=1$, does the Tsallis $q$-distribution (9) become the Boltzmann distribution, $f \sim \exp(-\beta H)$. For an ideal gas of one particle system, $H = p^2/2m$, from the above function, directly one can obtain the Maxwell $q$-distribution function, $f \sim [1-(1-q)\beta p^2/2m]^{1/1-q}$. The new form of Tsallis $q$-distribution (9) is a result based on the probabilistically independent postulate employed in nonextensive statistical mechanics. With this basic postulate, the $q$-entropy is *nonextensive* not only, but also is the energy [43]. However, quite questioningly, up to now nonextensive statistical mechanics develops without taking into consideration the *nonextensivity of energy*.

We now search for possible dynamical property of the new form of Tsallis $q$-distribution Eq.(9) from a general Fokker-Planck equation. Following the lines of previous work [16], we can assume the $q$-distribution Eq.(9) to be a stationary-state solution of the Fokker-Planck equation and then search for if it is a possible physical solution compatible with the dynamical functions in the Langevin equation of a dynamical system. If it is so, then the stationary-state solution can describe the long-times dynamical behavior of such a dynamical system.

We still starts with a general dynamical system of the two-variable Brownian motion of a particle, with mass $m$ and the Hamiltonian, $H = p^2/2m + \varphi(x)$, in an *arbitrary* potential $\varphi(x)$ (whether long- range or short-range force). The Langevin equations of the dynamical system are

$$\frac{dx}{dt} = \frac{p}{m}, \quad \frac{dp}{dt} = -\frac{d\varphi}{dx} - \zeta\frac{p}{m} + F_p(t), \tag{10}$$

where $\zeta$ is the frictional coefficient. The noise is Gaussian and it is delta-function correlated,

$$\langle F_p(t)F_p(t')\rangle = 2B\delta(t-t'). \tag{11}$$

Then the corresponding Fokker-Planck equation to the Langevin equations is given

for the noise-averaged distribution function [49, 16] by

$$\frac{\partial f}{\partial t} = -\frac{\partial}{\partial x}\frac{p}{m}f + \frac{\partial}{\partial p}\left(\frac{d\varphi}{dx} + \zeta\frac{p}{m}\right)f + B\frac{\partial^2 f}{\partial p^2}, \quad (12)$$

The stationary-state solution of this Fokker-Planck equation satisfy

$$-\frac{p}{m}\frac{\partial f}{\partial x} + \frac{d\varphi}{dx}\frac{\partial f}{\partial p} + \frac{\zeta}{m}(1 + p\frac{\partial}{\partial p})f + B\frac{\partial^2 f}{\partial p^2} = 0. \quad (13)$$

Equivalently, it can be written as

$$-\frac{p}{m}\frac{\partial f^{1-q}}{\partial x} + \frac{d\varphi}{dx}\frac{\partial f^{1-q}}{\partial p} + \frac{\zeta}{m}[1-q + p\frac{\partial}{\partial p}]f^{1-q} + (1-q)Bf^{-q}\frac{\partial^2 f}{\partial p^2} = 0. \quad (14)$$

According to Eq.(8) or Eq.(9), the new form of the Tsallis $q$-distribution for the above dynamical system is written as

$$f \sim [1-(1-q)\beta\varphi]^{\frac{1}{1-q}}\left[1-(1-q)\beta\frac{p^2}{2m}\right]^{\frac{1}{1-q}} \equiv R^{\frac{1}{1-q}}Q^{\frac{1}{1-q}}, \quad (15)$$

where one has denoted

$$R \equiv 1-(1-q)\beta\varphi, \quad Q \equiv 1-(1-q)\beta\frac{p^2}{2m}. \quad (16)$$

If Eq.(15) is a stationary-state solution of the Fokker-Planck equation, then put into Eq.(14), one has

$$\frac{Q}{m}\frac{d\beta\varphi}{dx}p + \frac{R}{2m^2}\frac{d\beta}{dx}p^3 - \frac{\beta R}{m}\frac{d\varphi}{dx}p + \frac{\zeta}{m}RQ - \frac{\zeta\beta}{m^2}Rp^2 + BRQ^{\frac{-q}{1-q}}\frac{\partial^2}{\partial p^2}Q^{\frac{1}{1-q}} = 0, \quad (17)$$

where for the last term one has

$$\frac{\partial^2}{\partial p^2}Q^{\frac{1}{1-q}} = -\frac{\beta}{m}Q^{\frac{1}{1-q}-1}\left(1-q\frac{\beta}{m}Q^{-1}p^2\right). \quad (18)$$

Then Eq.(17) becomes

$$m\left(R^{-1}\frac{d\beta\varphi}{dx}p + \zeta\right)Q^2 + \left(\frac{1}{2}\frac{d\beta}{dx}p^3 - \zeta\beta p^2 - m\beta\frac{d\varphi}{dx}p - mB\beta\right)Q + qB\beta^2 p^2 = 0. \quad (19)$$

Substituting $Q = [1-(1-q)\beta\frac{p^2}{2m}]$ into Eq.(19), one derives

$$\frac{1-q}{4m}\beta\left[(1-q)\beta R^{-1}\frac{d\beta\varphi}{dx} - \frac{d\beta}{dx}\right]p^5 + (1-q)(3-q)\frac{\beta^2}{4m}\zeta\ p^4$$

$$+\left[\frac{1}{2}\frac{d\beta}{dx}+(1-q)\frac{\beta^2}{2}\frac{d\varphi}{dx}-(1-q)\beta R^{-1}\frac{d\beta\varphi}{dx}\right]p^3+\left[B\beta^2-(2-q)\zeta\beta\right]p^2$$

$$+m\left(R^{-1}\frac{d\beta\varphi}{dx}-\beta\frac{d\varphi}{dx}\right)p+m(\zeta-B\beta)=0. \tag{20}$$

If $q=1$, we find $d\beta/dx=0$ and the fluctuation-dissipation theorem $\zeta=B\beta$, a physical situation for thermal equilibrium known well in Boltzmann-Gibbs statistics. If $q\neq 1$, very clearly, from Eq.(20) we can determine the following three identities to be satisfied for the Tsallis $q$-distribution (15). Namely, if the Tsallis distribution (15) is a stationary-state solution of the Fokker-Planck equation, then it must fulfill the three identities,

$$\frac{d\beta}{dx}=(1-q)\beta^2\frac{d\varphi}{dx}, \tag{21}$$

$$\zeta=0, \tag{22}$$

$$B=0. \tag{23}$$

The three identities can determine possible dynamics compatible with the Tsallis $q$-distribution function (15). As compared with the previous three identities [16], $d\beta/dx=0, \zeta=0,$ and $B=0$, obtained for the current form of Tsallis distribution, the above identity Eq.(21) stands for one nonequilibrium dynamical property of the system. Unsatisfactorily, we determine $\zeta=B=0$, do not obtain a generalized fluctuation-dissipation theorem as expected for the frictional coefficient $\zeta$ and the quantity $B$, e.g. $B\beta=(2-q)\zeta$, or other form.

Eqs.(21)-(234) represent one nonequilibrium dynamical property of the system with an arbitrary potential when it reaches to the nonequilibrium stationary-state, with no friction ($\zeta=0$) and irrelated noise ($B$=0). In this case, the corresponding Langevin dynamical equation (10) becomes

$$\frac{dp}{dt}=-\frac{d\varphi}{dx}+F_p(t), \tag{24}$$

where the noise is irrelated, $<F_p(t)F_p(t')>=0$ due to $B$=0. Eq.(24) is the dynamical equation for the system governed *only* by the potential field $\varphi$. The potential can be *arbitrary* one, whether long-range or short-range force. The new form of Tsallis distribution can describe the nonequilibrium dynamical property for such a system as governed by the Langevin equation (24). If the potential function is the gravitational one, it describes the dynamical property of the particles evolving in a self-gravitating system, where the gravitation is the only one force among the particles, e.g. the dark matter is now just thought of such a physical situation. In other words, Eq.(24) is Langevin equation for the dark matter. Our results show that the new form of Tsallis distribution (15) can be a stationary-state solution of the Fokker-Planck equation (12) under the situation (21)-(23) and supply one statistical description for the nonequilibrium dynamical property of the system characterized by Eq.(24), so to be one potential distribution function for the dark matter distribution.

The nonextensive parameter is now given exactly by the relation,

$$1-q = \frac{d\beta}{dx} \bigg/ \beta^2 \frac{d\varphi}{dx}. \tag{25}$$

The *nonextensivity* ($q \neq 1$) stands for a degree of deviation from the thermal equilibrium of the nonequilibrium dynamical system governed by an arbitrary potential field and thus has the clearly physical meaning. If take $d\beta/dx = 0$ or $dT/dx = 0$ (thermal equilibrium), one has $q = 1$ and the Tsallis $q$-distribution (16) becomes Boltzmann distribution. The reader might also be interested in some recent applications of nonextensive statistical mechanics, where a physical meaning of the parameter $q$ is introduced to astrophysics [17,19] and plasmas [20].

In the end, we would like to make remarks on the probabilistically independent problem. The probabilistic independence at the very start was as a basic postulate for nonextensive statistical mechanics [43,1-4]. Under this postulate, the $q$-entropy is *nonextensive*, satisfying Eq.(4), and nonextensive statistical mechanics is studied and developed. On the other hand, the probabilistically independent postulate also requires the energy to be nonextensive [43]. Namely, from the probabilistic independence one also can derive the relation for the energy $U$, composed of two probabilistically independent parts *a* and *b*,

$$U(a \oplus b) = U(a) + U(b) + (1-q)\beta\, U(a)U(b), \tag{26}$$

which appears to coexist with the relation for the *q*-entropy, $S(a \oplus b) = S(a) + S(b) + (1-q)k^{-1}S(a)S(b)$. Usually, the nonextensive statistics has been developed *only* by taking Eq.(4) for the *q*-entropy as the basic precondition but ignoring the coexisted Eq.(26) with it for the energy, leading to the current form of Tsallis distribution, Eq.(5) or Eq.(6). One postulate leads to two coexisted results. When the nonextensiv statistics selected one but discarded the other one, without interpretation, it had been incomplete theoretically.

In fact, from the second law of thermodynamics, e.g. $dU = TdS$ (if the volume is fixed), we may find that it is hard to image that the entropy is nonextensive but the energy is extensive. When we use the new form of the Tsallis distribution defined by Eq.(8) or Eq.(9), both the relation Eq.(26) for the energy's nonextensivity and the relation Eq.(4) for the entropy's nonextensivity has actually been taken into consideration.

In conclusion, we have expressed a new understanding for the *q*-exponential of a sum based on the probabilistically independent postulate in nonextensive statistical mechanics. Namely, the *q*-exponential of a sum can be applied as the product of the *q*-exponential by Eq.(8). Under this framework, we suggest a new form of Tsallis distribution (9), which incarnates the entropy's nonextensivity not only but the energy's nonextensivity. It is one reasonable scheme to solve the problems such as the current form of Tsallis distribution contravenes the basic postulate of the probabilistic independence, selects the entropy's nonextensivity but discards the coexisted energy's nonextensivity with it, and only stands for a simple isothermal or thermal equilibrium situation etc. The nonextensive parameter is exactly given by the relation Eq.(25) and so it has a clearly physical meaning. It ($q \neq 1$) stands for a degree of deviation from the thermal equilibrium of the dynamical system under an arbitrary potential field.

We show that the new Tsallis distribution (15) based on Eq.(8) can be a stationary-state solution of the Fokker-Planck equation (12). It is a physical solution with the three identities (21)-(23), so it can supply the statistical description for the nonequilibrium dynamics of the Hamiltonian systems governed by *any* potential when it reaches to the nonequilibrium stationary-state. If the potential is the gravitational one, it can describe the nonequilibrium dynamical property of particles evolving in a self-gravitating system, *e.g.* the dark matter is just such a physical situation. If the new Tsallis distribution were employed for the N-body simulation for a self-gravitating system, the results would be expected to be consistent with the dark matter halos.

**Reference**s